\documentclass[10pt,a4paper,twoside]{article}
\usepackage{epsfig}
\usepackage{baltlat6}
\usepackage{wrapfig}
\usepackage{array}
\usepackage{dcolumn}
\usepackage{longtable}

\begin{document}
\ \
\vspace{-0.5mm}

\setcounter{page}{95}
\vspace{-2mm}

\titlehead{Baltic Astronomy, vol.\,19, 95--110, 2010}

\titleb{CHEMICAL COMPOSITION OF THE RS CVn-TYPE STAR\\ LAMBDA
ANDROMEDAE}

\begin{authorl}
\authorb{G. Tautvai\v{s}ien\.{e}}{1},
\authorb{G. Barisevi\v{c}ius}{1},
\authorb{S. Berdyugina}{2},
\authorb{Y. Chorniy}{1} and\\
\authorb{I. Ilyin}{3}
\end{authorl}

\begin{addressl}
 \addressb{1}{Institute of Theoretical Physics and Astronomy, Vilnius
University,\\ Go\v{s}tauto 12, Vilnius LT-01108, Lithuania}

\addressb{2}{Kiepenheuer Institut f\"ur Sonnenphysik, Sch\"oneckstrasse
6, Freiburg D-79104,\\ Germany}

\addressb{3}{Astrophysical Institute Potsdam, An der Sternwarte 16,
Potsdam D-14482,\\ Germany} \end{addressl}

\submitb{Received: 2010 June 8; accepted: 2010 June 15}

\begin{summary} Photospheric parameters and chemical composition are
determined for the single-lined chromospherically active RS~CVn-type
star $\lambda$~And (HD~222107).  From the high resolution spectra
obtained on the Nordic Optical Telescope, abundances of 22 chemical
elements and isotopes, including such key elements as $^{12}{\rm C}$,
$^{13}{\rm C}$, N and O, were investigated.  The differential line
analysis with the MARCS model atmospheres gives $T_{\rm eff}$ = 4830~K,
log\,$g$ = 2.8, [Fe/H] = --0.53, [C/Fe] = 0.09, [N/Fe] = 0.35, [O/Fe] =
0.45, C/N = 2.21, $^{12}$C/$^{13}$C = 14.  The $^{12}$C/$^{13}$C ratio
for a star of the RS~CVn-type is determined for the first time, and its
low value gives a hint that extra-mixing processes may start acting in
low-mass chromospherically active stars below the bump of the luminosity
function of red giants.  \end{summary}

\begin{keywords} stars:  RS~CVn binaries, abundances -- stars:
individual ($\lambda$ And = HD~222107) \end{keywords}

\resthead{Chemical composition of the RS CVn-type star Lambda
Andromedae}{G.~Tautvai\v{s}ien\.{e}, G. Barisevi\v{c}ius,
S.~Berdyugina et al.}

\sectionb{1}{INTRODUCTION}

The RS CVn-type stars have been studied thoroughly since 1965 when their
peculiar light curves were detected (Rodon\'{o} 1965; Chisari \& Lacona
1965) and a new distinct class of binaries was named (Olivier 1974; Hall
1976).  RS CVn binary systems are typically composed of two
late-type chromospherically active fast-rotating stars, at least one of
which has already evolved off the main sequence (Hall 1976).  Tidal
forces between the close components make their rotational period to be
equal to the orbital period.  Similarly to other cool active stars, RS
CVn-type variables are remarkable due to large starspots, strong
chromospheric plages, coronal X-ray and microwave emissions, as well as
strong flares in the optical, radio and other spectral regions.  General
properties of RS CVn systems are comprehensively described by Montesinos
et al.\ (1988).  The photometric brightness variation analysis, Doppler
imaging and spectral line analysis of RS CVn stars indicate that
starspots may cover more than 20\% of their surfaces (Rodon\'{o} et al.\
1995; Berdyugina et al.\ 1998b, 2000; Jeffers 2005; Alekseev {\&}
Kozhevnikova 2005).  One of the largest filling factors (44\%) was
found for the RS CVn binary V841\,Cen (Strassmeier et al.\ 2008).

The photospheric abundances of chemical elements in RS CVn-type stars
are found to be peculiar and indicate a combined action of various
physical processes related to activity (e.g.,  Pallavicini et al.\ 1992,
Tautvai\v{s}ien\.{e} et al.\ 1992; Randich et al.\ 1993, 1994; Savanov
{\&} Berdyugina 1994; Berdyugina et al.\ 1998a, 1999; Katz et al.\ 2003;
Morel et al.\ 2003, 2004).

A new era of investigations of RS CVn stars has begun with the launch
of the {\it Extreme Ultraviolet Explorer}, {\it Chandra} and {\it
XMM-Newton} satellites and the study of element abundances in the
coronal plasma.  Highly active stars show a depletion of elements with a
low first ionization potential (FIP) (e.g., Fe, Mg, Si) relative to
high-FIP elements (e.g., C, N, O, Ne), whereas active binaries with
medium activity show either no the FIP feature or a possible solar-like
FIP effect; some studies show that only low-FIP elements are sensitive
to the activity level, while this is not the case for high-FIP elements
(Drake 1996, 2002; Jordan et al.\ 1998; Audard et al.\ 2001, 2003;
Bowyer et al.\ 2000; Sanz-Forcada et al.\ 2004 and references therein).
In order to interpret the apparent coronal abundance anomalies and true
abundance differences, a detailed comparison of photospheric abundances
is needed.

We have started a detailed study of the photospheric abundances in RS
CVn stars, and observed a sample of 28 such stars on the Nordic Optical
Telescope.  We consider that this study will be useful in addressing the
issues of photospheric and coronal abundance patterns and mixing
processes in these stars.  Our aim is to determine abundances of more
than 20 chemical elements, including $^{12}$C, $^{13}$C, N, O and other
mixing-sensitive species.  We plan to investigate correlations between
abundance alterations of chemical elements in star atmospheres and their
physical macro parameters, such as the speed of rotation and the
magnetic field.

In this paper we present results of the analysis on one of the
brightest RS CVn binaries, $\lambda$ And (HR\,8961, HD\,222107),
consisting of a G8 III-IV star ($V$\,$\approx$\,3.8 mag) and a
spectroscopically undetected secondary.  According to Donati et al.\
(1995), $\lambda$~And has a mass of $0.65^{+0.6}_{-0.3} M_{\odot}$, and
the unseen secondary component of the system is a low-mass main-sequence
star or a brown dwarf of mass $0.08\pm 0.02 M_{\odot}$.  Since the
primary dominates the emission, the photospheric abundances can be
relatively well determined, and thus the system is an ideal candidate
for studying coronal-to-photospheric abundance patterns.  According to
Savanov \& Berdyugina (1994), the mass of $\lambda$~And primary is
1.2~$M_{\odot}$.

Calder (1938) was the first to discover the photometric variability of
$\lambda$~And, its amplitude sometimes reaches 0.3 mag.  Six years
later, Walker (1944) showed that $\lambda$~And is an SB1 with almost
circular orbit of period 20.5212 d. It is atypical for a member of RS
CVn type stars, because it is considerably out of synchronism, its
rotational period being 54.3 d (Gondoin 2007).  In most RS CVn binaries
the rotational period of both components is equal, within a few percent,
to the orbital period of the system.  Thus $\lambda$~And, whose orbit is
very close to circular, is a puzzle for the theory of tidal friction
(Zahn 1977), which predicts that rotational synchronization in close
binaries should precede circulization of the orbit.

From more than 26 years of photometry, Mirtorabi et al.\ (2003) found
that the mean light level of $\lambda$~And varies between $\langle
V\rangle_{\rm min}$ = 3.89 mag and $\langle V\rangle_{\rm max}$ = 3.77
mag.  These observations have shown that $\lambda$~And exhibits
semi-regular cyclic light variations between 4--5~yr and $\sim14$~yr.
From these data, well-defined light minima were noticed in 1979/1980,
1991 and 1997, and well-defined light maxima in 1985/1986 and
1999/2000.  Our observations of $\lambda$~And have been done during its
light maximum in August of 1999.

For $\lambda$~And, the asymmetrical shape of the light curve requires
two spots at different longitudes and these spots must be 800--1050~K
cooler than the surrounding photosphere (Bopp \& Noah 1980; Poe \& Eaton
1985; Donati et al.\ 1995; Padmakar \& Pandey 1999; Frasca et al.\
2008).  According to O`Neal et al.\ (1998), the dark spot filling factor
for $\lambda$~And was 0.14 in late summer of 1993 and even 0.23 in
January of 1995.  According to Frasca et al.\ (2008), the star radius is
$R = 7.51 R_{\odot}$ and the inclination of the rotational axis with
respect to the line of sight is $i=67^{\circ}$.  Donati et al.  (1995)
give $v$~sin\,$i$ = 6.5~km\,s$^{-1}$.

The investigation of possible dependencies of spot parameters, such as
the temperature and filling factor, on global stellar parameters, such
as the effective temperature, gravity and activity level (rotation rate,
differential rotation, etc.), is very important to better understand the
physical mechanisms at work in the formation and evolution of RS CVn
stars.

The correlation of the temperature difference between the quiet
photosphere and spots, $\Delta T$, with the effective temperature has
been determined by Berdyugina (2005).  On the average, $\Delta T$ is
larger for the hotter stars, with the values from about 2000~K for
G-type stars to 200~K for M4 stars.  This behavior is displayed both for
giants and main-sequence stars.  It is interesting to investigate the
role of the surface gravity on $\Delta T$ by selecting stars of nearly
the same temperature.

\sectionb{2}{OBSERVATIONS AND THE METHOD OF ANALYSIS}

The spectra were obtained in August of 1999 on the 2.56~m Nordic Optical
Telescope using the SOFIN echelle spectrograph with the optical camera,
which provided a spectral resolving power of $R \approx 80\,000$, for
26 spectral orders, each of $\sim$\,40~\AA, in the spectral region
500--830~nm.  Reductions of the CCD images were made with the `4A'
software package (Ilyin 2000).  In image processing, the procedures of
bias subtraction, spike elimination, flat field correction, scattered
light subtraction and extraction of spectral orders were used.  The
continuum was defined by a number of narrow spectral regions, selected
to be free of lines.  In the spectra of $\lambda$ And, we selected 118
atomic lines for the measurement of equivalent widths and 19 lines for
the comparison with synthetic spectra.  The equivalent widths were
measured by fitting a Gaussian function with the `4A' software package
(Ilyin 2000).  The measured equivalent widths of the lines and their
parameters are presented in Table~1.

The spectra were analysed using a differential model atmosphere
technique.  The EQWIDTH and BSYN program packages, developed at the
Uppsala Astronomical Observatory, were used to carry out the
calculations of theoretical equivalent widths of lines and synthetic
spectra.  First, using the EQWIDTH program package we obtained solar
abundances which were later used for the differential determination of
abundances of the program stars.  The solar model atmosphere was taken
from the Uppsala set (Gustafsson et al.\ 2008) with $T_{\rm eff}$ =
5777~K, log\,$g=4.44$ and $v_{\rm t}=0.8$~km\,s$^{-1}$ determined from
the Fe\,{\sc i} lines.

{
\footnotesize
\noindent
\extrarowheight=-.5pt
\tabcolsep=7pt
\begin{longtable}{lcccccD..{3.3}D..{3.3}}
\multicolumn{8}{c}{\parbox[c]{120mm}{\baselineskip=9pt
{\smallbf\ \ Table 1.}{\small\ \ Atomic line data and the
measured equivalent widths.
In the head of the table $\chi_l$ is the
excitation potential of the lower level, log~$gf$ is the oscillator
strength, $\delta \Gamma_6$
 is the correction factor to the van-der-Waals damping constant,
$\Gamma_{\rm rad}$ is the radiation damping constant,
$EW_{\star}$ and $EW_{\odot}$ are the equivalent widths of lines in
$\lambda$ And and in the solar spectrum. \lstrut}}}\\
\tablerule
\noalign{\vskip0.5mm}
\multicolumn{1}{l}{Element} &
\multicolumn{1}{c}{$\lambda$~(\AA)} &
\multicolumn{1}{c}{$\chi_l$\,(eV)} &
\multicolumn{1}{c}{log\,$gf$} &
\multicolumn{1}{c}{$\delta \Gamma_6$} &
\multicolumn{1}{c}{$\Gamma_{\rm rad}$\,(s$^{-1}$)} &
\multicolumn{1}{c}{$EW_{\star}$~(m\AA)} &
\multicolumn{1}{c}{$EW_{\odot}$~(m\AA)} \\
\noalign{\vspace{0.5mm}}
\tablerule
\noalign{\vspace{0.5mm}}
\endfirsthead
\multicolumn{8}{l}{{\smallbf\ \ Table 1.}{\small\ Continued\lstrut}}\\
\tablerule
\noalign{\vskip0.5mm}
\multicolumn{1}{l}{Element} &
\multicolumn{1}{c}{$\lambda$~(\AA)} &
\multicolumn{1}{c}{$\chi_l$\,(eV)} &
\multicolumn{1}{c}{log\,$gf$} &
\multicolumn{1}{c}{$\delta \Gamma_6$} &
\multicolumn{1}{c}{$\Gamma_{\rm rad}$\,(s$^{-1}$)} &
\multicolumn{1}{c}{$EW_{\star}$~(m\AA)} &
\multicolumn{1}{c}{$EW_{\odot}$~(m\AA)} \\
\noalign{\vspace{0.5mm}}
\tablerule
\noalign{\vspace{0.5mm}}
\endhead
\endfoot
Si\,{\sc i} &    5793.08 &       4.93 & $     -1.98$ &       1.30 &   1.95e+08 &   44.3 &   43.5 \\
		 &    6131.85 &       5.61 & $     -1.72$ &       1.30 &   1.00e+05 &   22.6 &   25.7 \\
\tablerule
Ca\,{\sc i}   &    5867.57 &       2.93 & $     -1.59$ &       1.80 &   2.62e+08 &   49.0 &   23.9 \\
	 	 &    6455.60 &       2.52 & $     -1.42$ &       1.80 &   4.65e+07 &   93.0 &   56.7 \\
	  	 &    6798.47 &       2.71 & $     -2.49$ &       1.80 &   1.94e+07 &   22.5 &    6.3 \\
\tablerule
Sc\,{\sc ii} &    5526.81 &       1.77 & $$      0.25$$ &       2.50 &
2.16e+08 &  109.5 &   78.7\\
	   	&    6300.69 &       1.51 & $$     -1.93$$ &       2.50 &   2.31e+08 &   15.6 &    5.6\\
\tablerule
Ti\,{\sc i}  &    5648.58 &       2.49 & $     -0.35$ &       2.50 &   5.06e+07 &   44.8 &   10.1 \\
	 	 &    5662.16 &       2.32 & $     -0.14$ &       2.50 &   6.03e+07 &   71.7 &   20.8 \\
	  	 &    5716.45 &       2.30 & $     -0.82$ &       2.50 &   6.05e+07 &   31.2 &    5.8 \\
	  	 &    5739.48 &       2.25 & $     -0.69$ &       2.50 &   6.61e+07 &   41.1 &    8.1 \\
	  	 &    5899.30 &       1.05 & $     -1.19$ &       2.50 &   6.58e+07 &  103.4 &   28.8 \\
	  	 &    5903.31 &       1.07 & $     -2.10$ &       2.50 &   6.46e+06 &   43.2 &    4.6 \\
	  	 &    5941.76 &       1.05 & $     -1.55$ &       2.50 &   7.13e+07 &   80.1 &   15.1 \\
	  	 &    5953.17 &       1.89 & $     -0.33$ &       2.50 &   8.61e+06 &   90.1 &   31.3 \\
	  	 &    5965.83 &       1.88 & $     -0.46$ &       2.50 &   8.91e+06 &   83.6 &   25.5 \\
	  	 &    6064.63 &       1.05 & $     -1.86$ &       2.50 &   9.93e+06 &   59.5 &    8.1 \\
	  	 &    6098.66 &       3.06 & $     -0.14$ &       2.50 &   5.43e+07 &   22.0 &    5.3 \\
	  	 &    6121.00 &       1.88 & $     -1.45$ &       2.50 &   1.23e+08 &   28.4 &    3.5 \\
	  	 &    6126.22 &       1.07 & $     -1.37$ &       2.50 &   9.93e+06 &   86.1 &   21.0 \\
	  	 &    6220.49 &       2.68 & $     -0.29$ &       2.50 &   4.61e+07 &   39.5 &    8.4 \\
	  	 &    6303.77 &       1.44 & $     -1.54$ &       2.50 &   1.75e+08 &   51.8 &    7.4 \\
	  	 &    6599.11 &       0.90 & $     -2.05$ &       2.50 &   1.22e+06 &   66.9 &    8.0 \\
\tablerule
V\,{\sc i}	&    5604.96 &       1.04 & $     -1.17$ &       2.50 &   8.47e+07 &   33.5 &    3.6 \\
	 &    5646.11 &       1.05 & $     -1.10$ &       2.50 &   8.53e+07 &   36.4 &    4.1 \\
	 &    5657.45 &       1.06 & $     -1.01$ &       2.50 &   8.47e+07 &   45.7 &    5.0 \\
	 &    5668.37 &       1.08 & $     -0.93$ &       2.50 &   8.39e+07 &   43.5 &    5.7 \\
	 &    5743.43 &       1.08 & $     -0.88$ &       2.50 &   7.67e+07 &   51.8 &    6.3 \\
	 &    6058.18 &       1.04 & $     -1.36$ &       2.50 &   3.94e+07 &   29.9 &    2.5 \\
	 &    6111.65 &       1.04 & $     -0.68$ &       2.50 &   3.90e+07 &   71.5 &   11.0 \\
	 &    6119.53 &       1.06 & $     -0.41$ &       2.50 &   3.94e+07 &   82.5 &   18.0 \\
	 &    6135.37 &       1.05 & $     -0.73$ &       2.50 &   3.90e+07 &   63.5 &    9.7 \\
	 &    6224.50 &       0.29 & $     -1.79$ &       2.50 &   1.22e+06 &   60.8 &    5.0 \\
	 &    6233.19 &       0.28 & $     -1.90$ &       2.50 &   1.35e+06 &   53.8 &    4.0 \\
	 &    6266.30 &       0.28 & $     -2.08$ &       2.50 &   1.37e+06 &   43.2 &    2.6 \\
	 &    6274.66 &       0.27 & $     -1.64$ &       2.50 &   3.10e+06 &   61.9 &    7.3 \\
	 &    6285.16 &       0.28 & $     -1.53$ &       2.50 &   2.91e+06 &   71.9 &    8.9 \\
\tablerule
Cr\,{\sc i}	&    5712.78 &       3.00 & $     -1.11$ &       2.50 &   1.78e+08 &   26.4 &   13.9 \\
	 &    5783.87 &       3.32 & $     -0.21$ &       2.50 &   9.98e+07 &   62.8 &   39.0 \\
	 &    5784.97 &       3.32 & $     -0.39$ &       2.50 &   9.98e+07 &   46.7 &   29.9 \\
	 &    5787.92 &       3.32 & $     -0.06$ &       2.50 &   1.01e+08 &   62.8 &   47.2 \\
	 &    6979.80 &       3.46 & $     -0.22$ &       2.50 &   2.53e+08 &   54.8 &   35.0 \\
\tablerule
Fe\,{\sc i}  &    5406.78 &       4.37 & $     -1.46$ &       1.40 &   1.77e+08 &   43.1 &   38.5 \\
	  	 &    5522.45 &       4.21 & $     -1.52$ &       1.40 &   8.97e+07 &   51.4 &   43.1 \\
	  	 &    5577.03 &       5.03 & $     -1.54$ &       1.40 &   6.89e+08 &   11.6 &   12.1 \\
	  	 &    5651.48 &       4.47 & $     -1.87$ &       1.40 &   1.63e+08 &   21.4 &   17.4 \\
	  	 &    5652.33 &       4.26 & $     -1.83$ &       1.40 &   8.18e+07 &   31.0 &   26.2 \\
	  	 &    5653.86 &       4.39 & $     -1.49$ &       1.40 &   1.79e+08 &   42.9 &   36.9 \\
      	 &    5679.03 &       4.65 & $     -0.75$ &       1.40 &   1.40e+08 &   66.8 &   64.6 \\
  	 &    5720.90 &       4.55 & $     -1.88$ &       1.40 &   1.92e+08 &   18.7 &   15.0 \\
Fe\,{\sc i}    	 &    5732.30 &       4.99 & $     -1.54$ &       1.40 &   6.89e+08 &   11.8 &   13.1 \\
	  	 &    5738.24 &       4.22 & $     -2.29$ &       1.40 &   2.01e+08 &   17.2 &   12.2 \\
	  	 &    5741.86 &       4.26 & $     -1.73$ &       1.40 &   2.10e+08 &   37.7 &   31.2 \\
	  	 &    5784.67 &       3.40 & $     -2.69$ &       1.40 &   7.53e+07 &   41.0 &   24.3 \\
	  	 &    5793.92 &       4.22 & $     -1.69$ &       1.40 &   2.11e+08 &   40.9 &   35.0 \\
	  	 &    5806.73 &       4.61 & $     -0.96$ &       1.40 &   1.86e+08 &   57.2 &   54.8 \\
	  	 &    5809.22 &       3.88 & $     -1.73$ &       1.40 &   5.06e+07 &   61.9 &   48.6 \\
	  	 &    5811.92 &       4.14 & $     -2.45$ &       1.40 &   3.76e+07 &   12.9 &   10.3 \\
	  	 &    5814.82 &       4.28 & $     -1.90$ &       1.40 &   2.11e+08 &   26.7 &   22.8 \\
	  	 &    6027.06 &       4.07 & $     -1.23$ &       1.40 &   8.85e+07 &   78.9 &   66.4 \\
	  	 &    6034.04 &       4.31 & $     -2.43$ &       1.40 &   1.58e+08 &   10.7 &    7.9 \\
	  	 &    6035.35 &       4.29 & $     -2.60$ &       1.40 &   6.92e+07 &    7.3 &    5.8 \\
	  	 &    6054.07 &       4.37 & $     -2.29$ &       1.40 &   1.66e+08 &   12.2 &    9.4 \\
	  	 &    6056.01 &       4.73 & $     -0.47$ &       1.40 &   1.85e+08 &   74.6 &   78.8 \\
	  	 &    6098.25 &       4.56 & $     -1.88$ &       1.40 &   2.62e+08 &   15.7 &   15.1 \\
	  	 &    6105.13 &       4.55 & $     -2.09$ &       1.40 &   1.94e+08 &   12.1 &   10.2 \\
	  	 &    6120.24 &       0.91 & $     -5.91$ &       1.40 &   2.71e+04 &   27.2 &    5.0 \\
	  	 &    6187.99 &       3.94 & $     -1.68$ &       1.40 &   4.60e+07 &   61.0 &   49.9 \\
	  	 &    6226.74 &       3.88 & $     -2.16$ &       1.40 &   5.42e+07 &   36.7 &   28.8 \\
	  	 &    6229.23 &       2.84 & $     -3.04$ &       1.40 &   1.45e+08 &   59.4 &   33.6 \\
	  	 &    6270.23 &       2.86 & $     -2.64$ &       1.40 &   1.45e+08 &   80.2 &   53.2 \\
	  	 &    6380.75 &       4.19 & $     -1.39$ &       1.40 &   7.35e+07 &   61.0 &   53.6 \\
	  	 &    6392.54 &       2.28 & $     -4.05$ &       1.40 &   1.65e+08 &   37.1 &   15.4 \\
	  	 &    6574.21 &       0.99 & $     -5.03$ &       1.40 &   3.38e+04 &   76.4 &   26.2 \\
	 	 &    6581.21 &       1.48 & $     -4.75$ &       1.40 &   1.56e+07 &   57.6 &   18.4 \\
	  	 &    6646.97 &       2.61 & $     -3.97$ &       1.40 &   7.57e+07 &   21.9 &    9.8 \\
	  	 &    6786.86 &       4.19 & $     -1.99$ &       1.40 &   1.99e+08 &   27.6 &   24.6 \\
	  	 &    6793.27 &       4.07 & $     -2.48$ &       1.40 &   7.08e+07 &   18.0 &   12.1 \\
	  	 &    6839.83 &       2.56 & $     -3.43$ &       1.40 &   9.31e+07 &   55.9 &   29.8 \\
	  	 &    6842.69 &       4.64 & $     -1.25$ &       1.40 &   2.23e+08 &   40.9 &   40.1 \\
	  	 &    6843.65 &       4.55 & $     -0.90$ &       1.40 &   1.92e+08 &   67.0 &   64.3 \\
	  	 &    6851.64 &       1.61 & $     -5.39$ &       1.40 &   1.46e+07 &   15.9 &    3.9 \\
	  	 &    6857.25 &       4.07 & $     -2.17$ &       1.40 &   2.53e+07 &   26.5 &   22.3 \\
	  	 &    6858.15 &       4.61 & $     -1.01$ &       1.40 &   1.92e+08 &   56.1 &   55.2 \\
	  	 &    6862.49 &       4.56 & $     -1.52$ &       1.40 &   3.54e+08 &   33.3 &   30.7 \\
	  	 &    7461.53 &       2.56 & $     -3.57$ &       1.40 &   1.21e+08 &   55.6 &   25.5 \\
\tablerule
Fe\,{\sc ii}    &    5256.93 &       2.88 & $     -4.25$ &       2.50 &   3.41e+08 &   21.3 &   19.9 \\
	  	 &    5264.81 &       3.22 & $     -3.26$ &       2.50 &   4.11e+08 &   40.2 &   47.6 \\
	  	 &    5534.84 &       3.24 & $     -2.98$ &       2.50 &   2.99e+08 &   56.6 &   60.0 \\
	  	 &    6113.33 &       3.22 & $     -4.26$ &       2.50 &   3.41e+08 &    9.9 &   12.0 \\
	  	 &    6369.46 &       2.89 & $     -4.28$ &       2.50 &   2.90e+08 &   15.9 &   20.1 \\
	  	 &    6383.71 &       5.55 & $     -2.25$ &       2.50 &   4.09e+08 &    5.1 &   10.9 \\
	  	 &    6456.39 &       3.90 & $     -2.28$ &       2.50 &   3.37e+08 &   52.3 &   66.1 \\
	  	 &    7711.72 &       3.90 & $     -2.72$ &       2.50 &   4.12e+08 &   34.3 &   49.8 \\
\tablerule
Co\,{\sc i}	&    5647.23 &       2.28 & $     -1.51$ &       2.50 &   1.66e+08 &   37.6 &   13.3 \\
	 &    6595.86 &       3.71 & $     -0.64$ &       2.50 &   6.84e+07 &   10.7 &    5.0 \\
	 &    6678.82 &       1.96 & $     -2.18$ &       2.50 &   2.01e+07 &   22.5 &    6.8 \\
\tablerule
Ni\,{\sc i}	&    5578.73 &       1.68 & $     -2.56$ &       2.50 &   5.43e+07 &   90.7 &   55.0 \\
	 &    5587.87 &       1.93 & $     -2.28$ &       2.50 &   1.25e+08 &   86.0 &   56.9 \\
	 &    5589.37 &       3.90 & $     -1.03$ &       2.50 &   1.30e+08 &   31.6 &   27.9 \\
         &    5593.75 &       3.90 & $     -0.67$ &       2.50 &   1.32e+08 &   46.4 &   45.2 \\
	 &    5643.09 &       4.16 & $     -1.13$ &       2.50 &   9.14e+07 &   15.6 &   15.3 \\
	 &    5748.35 &       1.68 & $     -3.12$ &       2.50 &   4.31e+07 &   59.3 &   29.0 \\
	 &    5805.22 &       4.17 & $     -0.48$ &       2.50 &   2.17e+08 &   41.5 &   43.0 \\
	 &    6111.08 &       4.09 & $     -0.72$ &       2.50 &   1.46e+08 &   36.4 &   35.5 \\
	 &    6128.98 &       1.68 & $     -3.27$ &       2.50 &   1.21e+07 &   56.8 &   23.6 \\
Ni\,{\sc i}         &    6130.14 &       4.26 & $     -0.89$ &       2.50 &   2.78e+08 &   20.0 &   21.0 \\
	 &    6378.25 &       4.15 & $     -0.76$ &       2.50 &   2.08e+08 &   35.2 &   31.5 \\
	 &    6598.60 &       4.23 & $     -0.85$ &       2.50 &   1.92e+08 &   26.7 &   24.5 \\
	 &    6635.13 &       4.42 & $     -0.68$ &       2.50 &   1.51e+08 &   22.4 &   24.7 \\
	 &    6772.32 &       3.66 & $     -0.86$ &       2.50 &   1.50e+08 &   56.5 &   50.9 \\
	 &    6842.03 &       3.66 & $     -1.36$ &       2.50 &   1.51e+08 &   33.0 &   26.1 \\
	 &    7001.55 &       1.93 & $     -3.48$ &       2.50 &   4.86e+07 &   30.4 &   11.1 \\
	 &    7715.59 &       3.70 & $     -0.87$ &       2.50 &   5.87e+07 &   58.5 &   52.0 \\
\tablerule
\end{longtable}
}

The atomic oscillator strengths and solar equivalent widths for the
lines (Table~1) were taken from Gurtovenko \& Kostik (1989).  The Vienna
Atomic Line Data Base (VALD) (Piskunov et al.\ 1995) was used in
preparing other input data for the calculations.  In addition to
thermal and microturbulent Doppler broadening of lines, atomic line
broadening by radiation damping and van der Waals damping were
considered in the calculation of abundances.  In most cases the hydrogen
pressure damping of metal lines was treated using the modern quantum
mechanical calculations by Anstee \& O'Mara (1995), Barklem \& O'Mara
(1997) and Barklem et al.\ (1998).  When using the Uns\"{o}ld (1955)
approximation, correction factors to the classical van der Waals damping
approximation by widths $(\Gamma_6)$ were taken from Simmons \&
Blackwell (1982).  For all other species a correction factor of 2.5 was
applied to the classical $\Gamma_6$ $(\Delta {\rm log}C_{6}=+1.0$),
following M\"{a}ckle et al.\ (1975).

\subsectionb{2.1}{Atmospheric parameters}

Initially, the effective temperature $T_{\rm eff}$ of $\lambda$~And was
derived and averaged from the intrinsic color indices $(B-V)_0$ and
$(b-y)_0$ using corresponding calibrations by Alonso et al.\ (1999).
The color indices $B$--$V$ = 0.984 and $b$--$y$ = 0.625 were taken from
van Leeuwen et al.\ (2007) and Hauck \& Mermilliod (1998), respectively.
A small dereddening correction of $E_{B-V}$ = 0.01, estimated by the
Hakkila et al.\ (1997) software, was taken into account.

The agreement between the temperatures deduced from the two color
indices was quite good, the difference was only 20~K.  No obvious trend
of the Fe\,{\sc i} abundances with the excitation potential was found
(Figure 1).  The surface gravity log\,$g$ was found by adjusting the
model gravity to yield the same iron abundance from the Fe\,{\sc i} and
Fe\,{\sc ii} lines.  Microturbulent velocity $v_{\rm t}$ value
corresponding to minimal line-to-line Fe\,{\sc i} abundance scattering
was chosen as the correct value.  Consequently, the [Fe/H] values do not
depend on the equivalent widths of lines (Figure 2).

\begin{figure}[!th]
\vbox{
\centerline{\psfig{figure=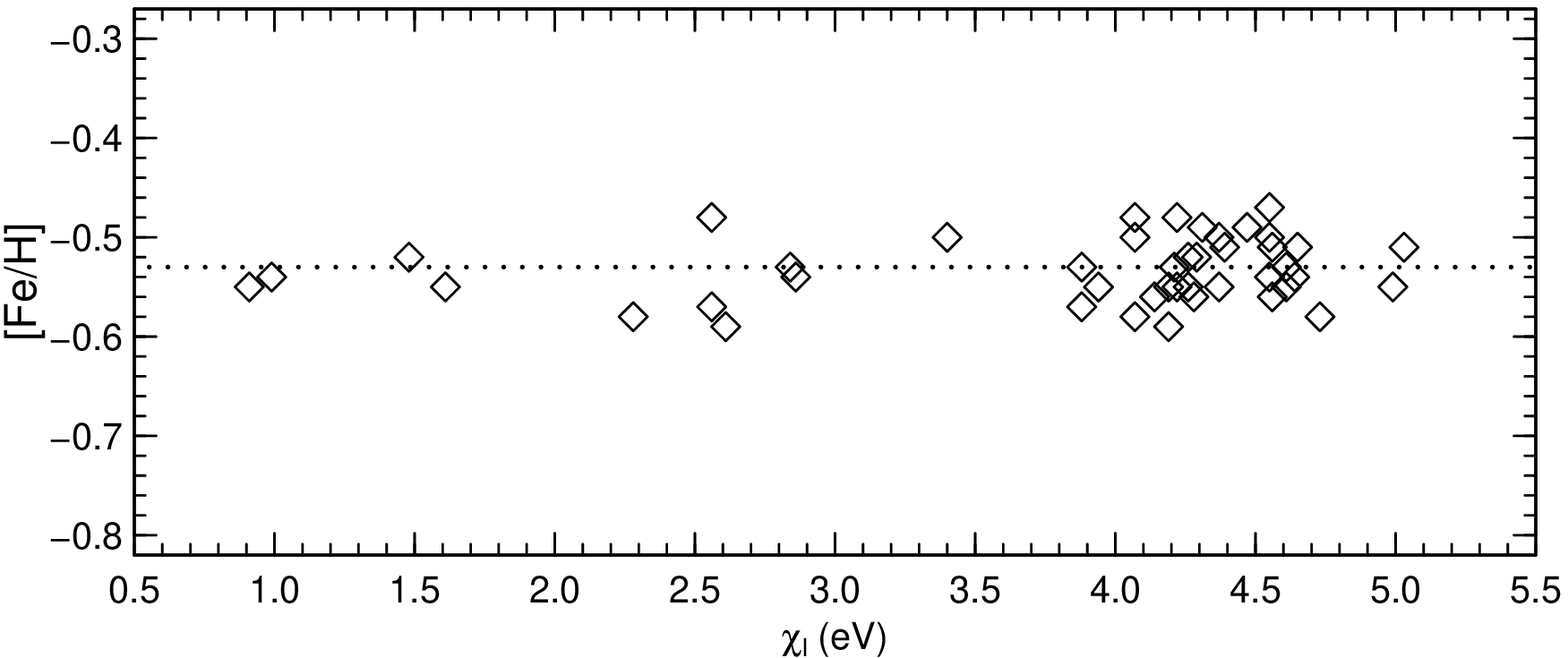,width=120truemm,angle=0,clip=}}
\captionb{1}{The [Fe\,{\sc i}/H] abundance values versus the lower
excitation potential  $\chi_{\rm exc}$ for $\lambda$~And. The mean
abundance ([Fe\,{\sc i}/H]$=-0.53$~dex) is shown by a dotted line.}
}
\end{figure}
\vspace{3mm}

\begin{figure}[!th]
\vbox{
\centerline{\psfig{figure=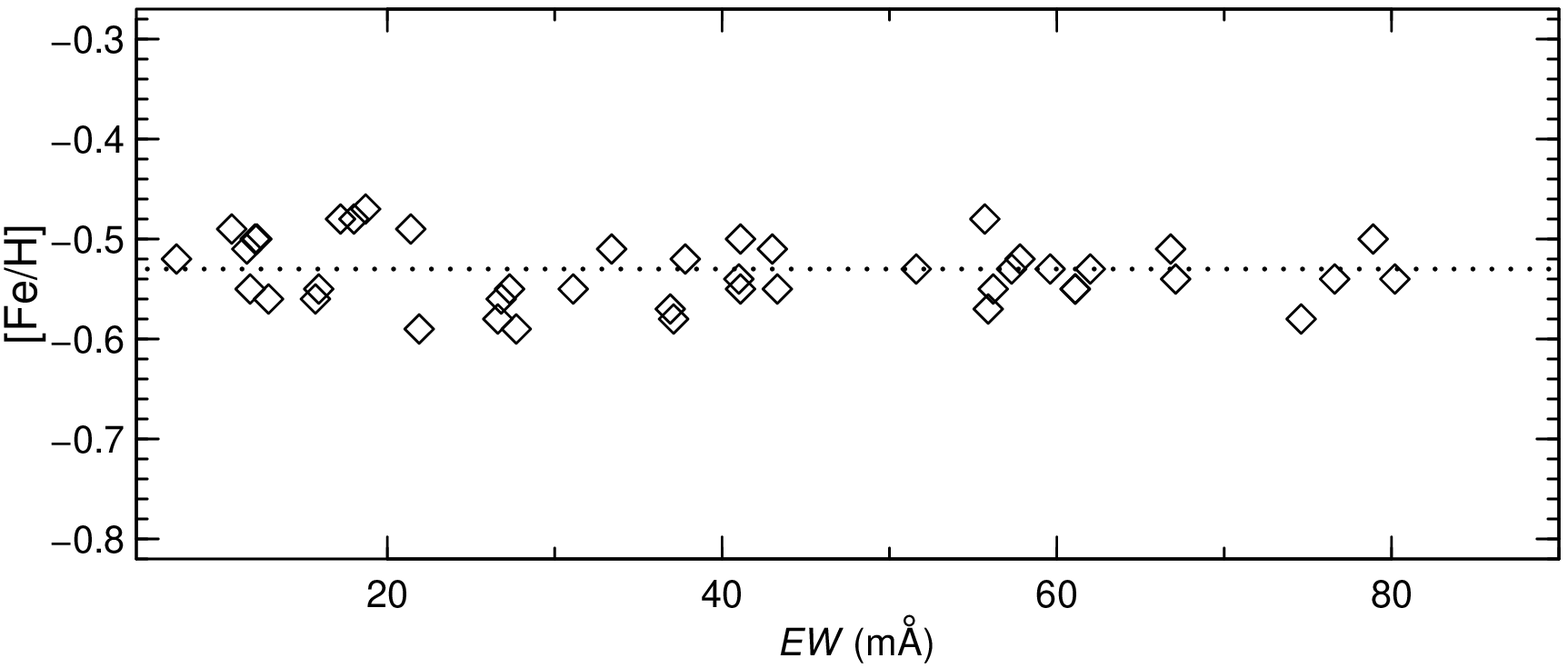,width=120truemm,angle=0,clip=}}
\captionb{2}{[The Fe\,{\sc i}/H] abundance values versus the
equivalent widths for $\lambda$~And.  The mean abundance ([Fe\,{\sc
i}/H]$=-0.53$~dex) is shown by a dotted line.}
}
\end{figure}

\subsectionb{2.2}{Mass determination}

The mass of the $\lambda$ And was evaluated from its effective
temperature, luminosity and the Girardi et al.\ (2000) isochrones.  The
luminosity ${\rm log}(L/L_{\odot})=1.37$ was calculated from the
Hipparcos parallax $\pi$ = 37.87~mas (van Leeuwen 2007), $\langle
V\rangle_{\rm max}$ = 3.77 mag (Mirtorabi et al.\ 2003), the bolometric
correction calculated according to Alonso et al.\ (1999) and $E_{B-V}$ =
0.01.  The mass of $\lambda$~And is found to be 1.1\,$M_{\odot}$.  This
value is close to the mass 1.2\,$M_{\odot}$ determined by Savanov \&
Berdyugina (1994).

\subsectionb{2.3}{Synthetic spectra}

The method of synthetic spectra was used to determine the carbon
abundance from the C$_2$ line at 5135.5~{\AA} using the
the Gonzalez et al.\ (1998) molecular data.

The interval 798--813~{\AA}, containing strong $^{12}{\rm C}^{14}{\rm
N}$ and $^{13}{\rm C}^{14}{\rm N}$ features, was used for determining
the nitrogen abundance and the $^{12}{\rm C}/^{13}{\rm C}$ ratio.  The
molecular data for this band were provided by Bertrand Plez (University
of Montpellier II).  The $^{12}{\rm C}/^{13}{\rm C}$ ratio was
determined from (2,0) $^{13}{\rm C}^{12}{\rm N}$ feature at
8004.7~{\AA}.  All log\,$gf$ values were calibrated to fit to the solar
spectrum by Kurucz (2005) with solar abundances from Grevesse \& Sauval
(2000).  In Figure~3 we show several examples of synthetic spectra in
the vicinity of $^{12}{\rm C}^{14}{\rm N}$ lines.
\vskip2mm

\vbox{
\centerline{\psfig{figure=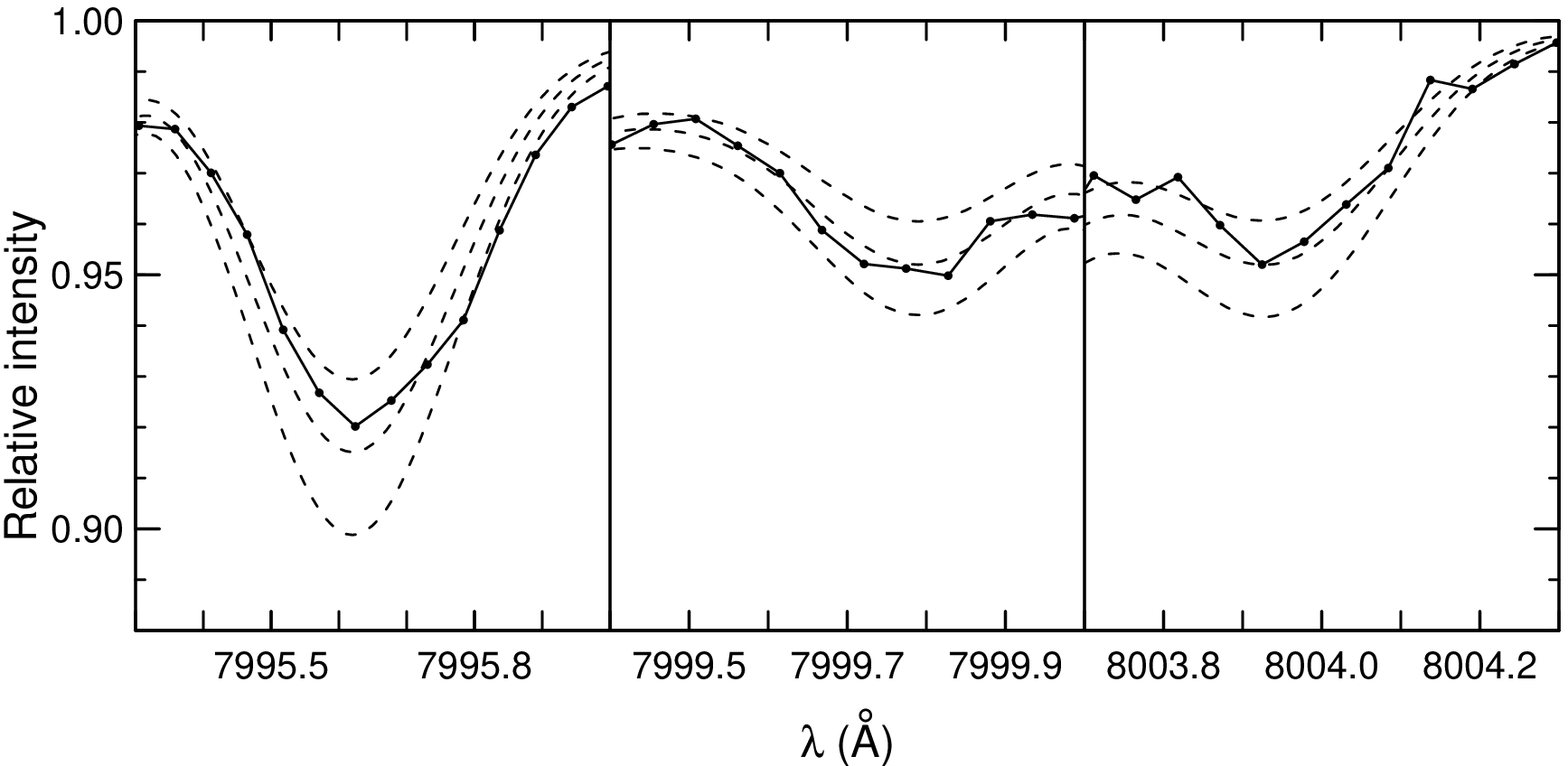,width=124truemm,angle=0,clip=}}
\captionb{3}{Synthetic spectrum fits to the three \textsuperscript{12}CN
lines for $\lambda$~And. The observed spectra
are shown as solid lines joining the dots. The dashed lines
denote the synthetic spectra with [N/Fe] = 0.25, 0.35 and 0.45~dex
downward.}
}
\vspace{5mm}

The oxygen abundance was determined from the forbidden [O\,{\sc i}] line
at 6300.31~\AA\ (Figure~4) with the oscillator strengths for
\textsuperscript{58}Ni and \textsuperscript{60}Ni from Johansson et al.
(2003), the log~$gf = -9.917$ value, obtained by fitting to the solar
spectrum of Kurucz (2005) and log~$A_{\odot}=8.83$ taken from Grevesse
\& Sauval (2000).

The abundance of Na\,{\sc i} was estimated using the line 5148.84~\AA\
which due to rotational broadening is blended by the Ni\,{\sc i} line at
5148.66~\AA.  These two lines are distinct in the Sun, so we were able
to calibrate their log\,$gf$ values using the solar spectrum.  However,
the sodium abundance value in $\lambda$ And is affected by uncertainty
of nickel abundance determination, originating from the Equivalent
Widths method.  Fortunately, the line-to-line scatter of [Ni/H]
determinations from 19 lines of Ni\,{\sc i} was as small as 0.04~dex.

For the evaluation of Zr\,{\sc i} abundance the lines at 5385.13~\AA,
6127.48~\AA\ and 6134.57~\AA\ were used.  Evaluation of Y\,{\sc ii}
abundance (Figure~5) was based on 5402.78~\AA, Pr\,{\sc ii} on
5259.72~\AA, La\,{\sc ii} on 6390.48~\AA, Ce\,{\sc ii} on 5274.22~\AA\
and 6043.38~\AA, and Nd\,{\sc ii} on 5276.86~\AA\ lines.

\begin{figure}[!th]
\vbox{
\centerline{\psfig{figure=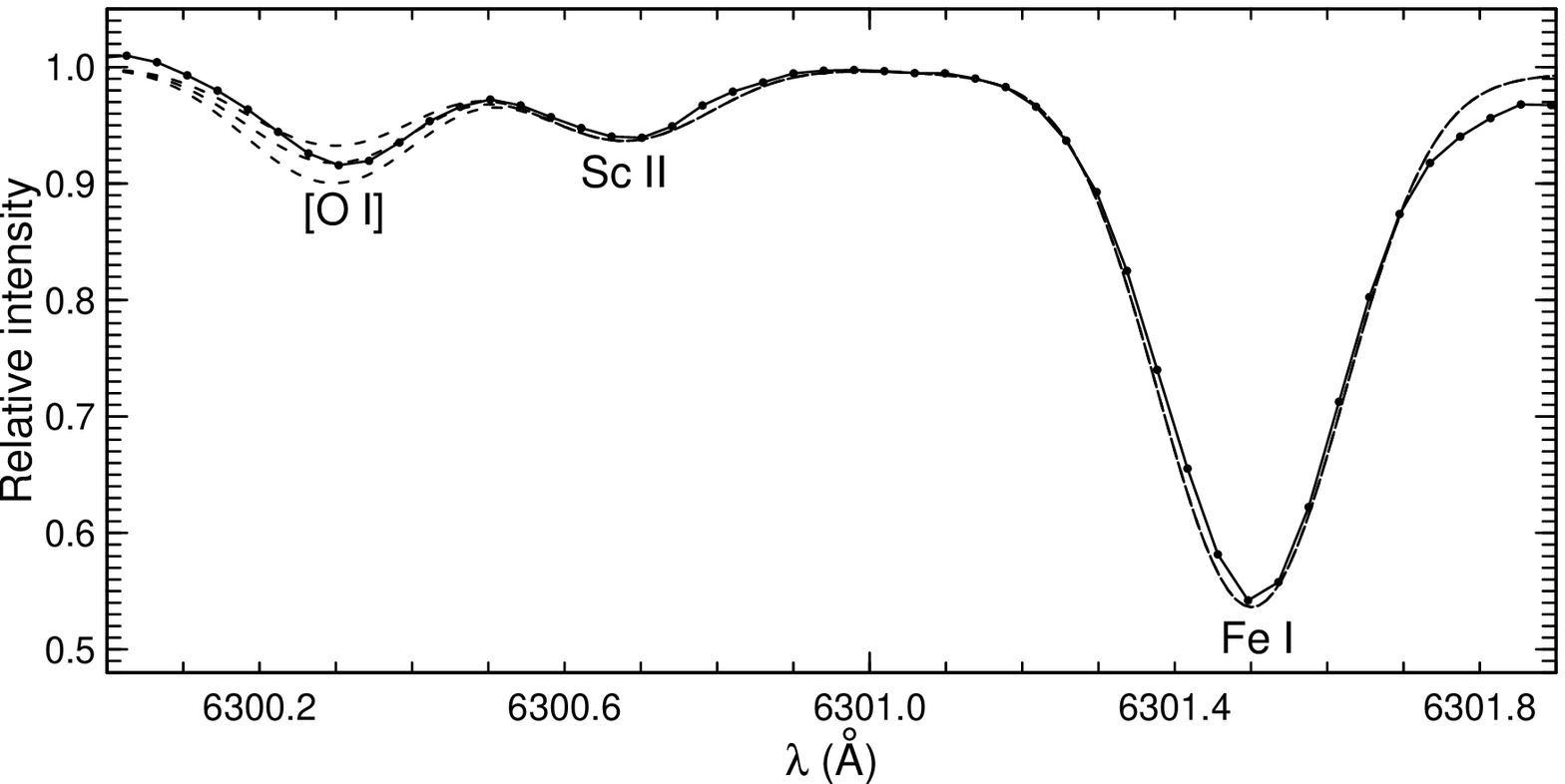,width=124truemm,angle=0,clip=}}
\captionb{4}{Synthetic spectrum fit to the forbidden [O\,{\sc i}] line
at 6300~\AA\ for $\lambda$~And.  The observed spectrum is shown as
the solid line joining the dots.  The dashed lines denote the
synthetic spectra with [O/Fe] = 0.35, 0.45 and 0.55 dex downward.}
}
\end{figure}
\vspace{3mm}

\begin{figure}[!th]
\vbox{
\centerline{\psfig{figure=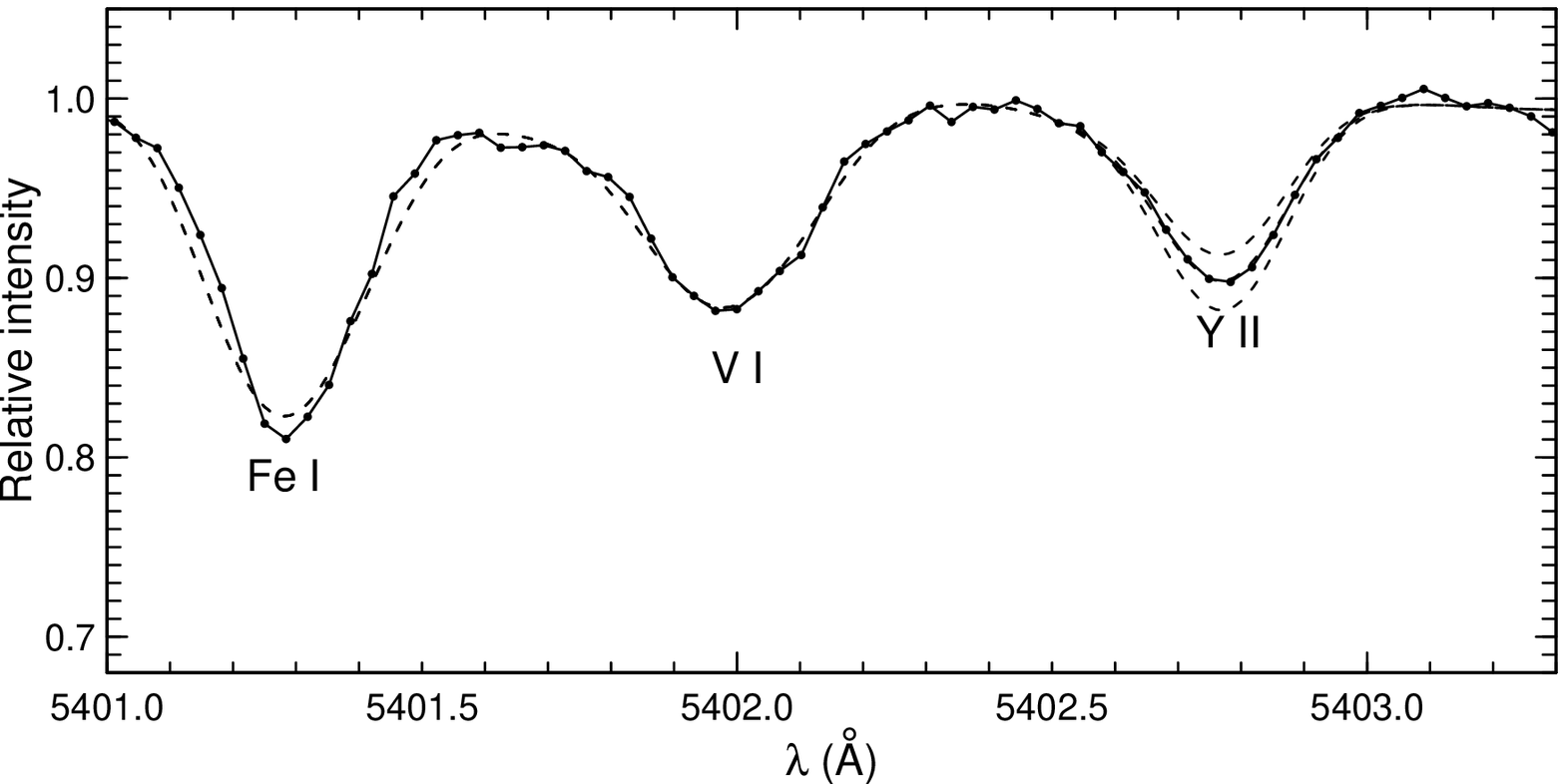,width=124truemm,angle=0,clip=}}
\captionb{5}{Synthetic spectrum fit to the Y\,{\sc ii} line at
5402.78~\AA\ for $\lambda$~And.  The observed spectrum is shown as the
solid line joining the dots.  The dashed lines denote the
synthetic spectra with [Y/Fe] = 0.05, 0.15 and 0.25 dex downward.}
}
\vspace{3mm}
\end{figure}

The abundance of Eu\,{\sc ii} was determined from the 6645.10~\AA\ line
(Figure~6).  The hyperfine structure of Eu\,{\sc ii} was taken into
account when calculating the synthetic spectrum.  The wavelength,
excitation energy and total log~$gf = 0.12$ were taken from Lawler et
al.\ (2001), the isotopic meteoritic fractions of $^{151}{\rm Eu}$,
47.77\%, and $^{153}{\rm Eu}$, 52.23\%, and isotopic shifts were taken
from Biehl (1976).

Due to the rotation of RS CVn stars, their lines are broadened, so it is
important to use a correct value of $v\,{\rm sin}\,i$ in the synthetic
spectrum production.  We used the most recent value, $v\,{\rm
sin}\,i=6.9~{\rm km\,s}^{-1}$, taken from De~Medeiros et al.\ (2002),
which fits our data quite well.  Previous authors provided higher values
of $v\,{\rm sin}\,i$, up to $20~{\rm km\,s}^{-1}$ (Uesugi \& Fukuda
1982), which seem to be overestimated.

\vskip2mm

\begin{figure}[!th]
\vbox{
\centerline{\psfig{figure=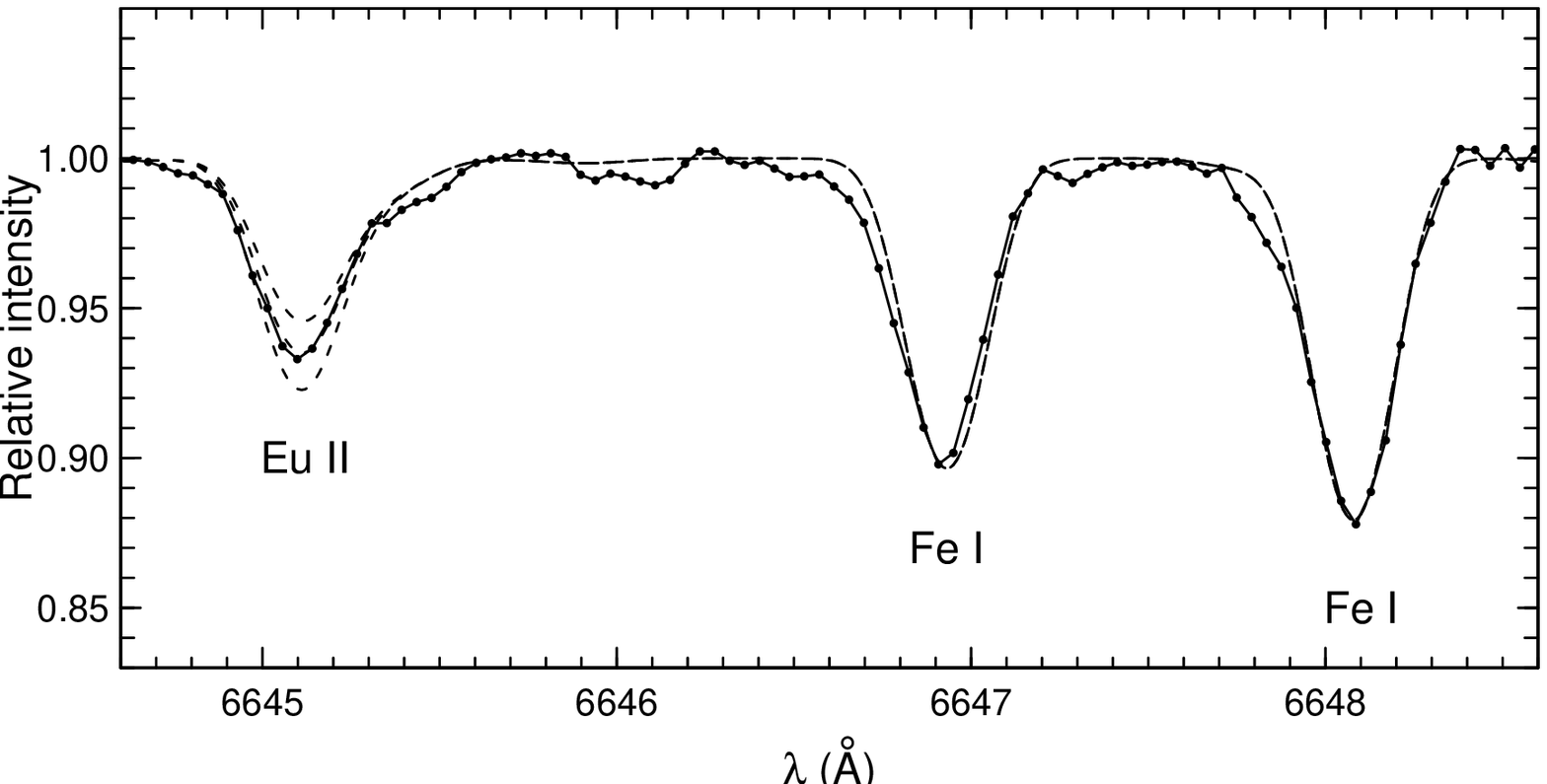,width=124truemm,angle=0,clip=}}
\captionb{6}{Synthetic spectrum fit to the Eu\,{\sc ii} line at
6645.10~\AA\ for $\lambda$~And.  The observed spectrum is shown as the
solid line joining the dots.  The dashed lines denote the synthetic
spectra with ${\rm [Eu/Fe]}$ = 0.28, 0.38 and 0.48~dex downward.}
}
\end{figure}

\subsectionb{2.4}{Estimation of uncertainties}

Sources of errors in our research can be divided into two distinct
categories.  The first one includes the errors that affect only single
lines (e.g., the errors in equivalent width measurements or line
parameters).  Other sources of observational errors, such as the
continuum placement or background subtraction problems, also are partly
included in the equivalent width errors.  The scatter of the deduced
line abundances, $\sigma$, presented in Table~3, gives an estimation of
the uncertainty due to random errors in the line parameters (the mean
value of $\sigma$ is 0.05 dex).  Thus the uncertainties in the derived
abundances that are the result of random errors are close to this value.

The second category includes the errors which affect all the lines
together.  These are mainly the model errors, such as errors in the
effective temperature, surface gravity, microturbulent velocity, etc.).
The sensitivity of the abundance estimates to changes in the atmospheric
parameters by the assumed errors is illustrated in Table~2.

Since the abundances of C, N and O are bound together by the molecular
equilibrium in the stellar atmosphere, we have also investigated how the
error in one of them typically affects the abundance determination of
another.  $\Delta{\rm [O/H]}=0.10$ causes $\Delta{\rm [C/H]}=0.04$ and
$\Delta{\rm [N/H]}=0.03$; $\Delta{\rm [C/H]}=0.10$ causes $\Delta{\rm
[N/H]}=-0.12$ and $\Delta{\rm [O/H]}=0.03$.  $\Delta {\rm [N/H]}=0.10$
has no effect on either the carbon or the oxygen abundances.

\begin{table}

\begin{center}
\vbox{\small
\tabcolsep=5pt
\begin{footnotesize}
\begin{tabular}{lD..{3.3}D..{3.3}D..{3.3}|lD..{3.3}D..{3.3}D..{3.3}}
\multicolumn{8}{c}{\parbox{115mm}{
  {\small \bf \ \ Table 2.~}{\small The sensitivity of
abundances to the changes in the atmospheric parameters.
The table entries  show the effects on the logarithmic abundance
relative to hydrogen, $\Delta$\,[A/H].
}}}\\
\tablerule
\multicolumn{1}{l}{Element}&
\multicolumn{1}{c}{$\Delta T_{\rm eff}$}&
\multicolumn{1}{c}{$\Delta {\rm log}~g$}&
\multicolumn{1}{c|}{$\Delta v_{\rm t}$}&
\multicolumn{1}{l}{Element}&
\multicolumn{1}{c}{$\Delta T_{\rm eff}$}&
\multicolumn{1}{c}{$\Delta {\rm log}~g$}&
\multicolumn{1}{c}{$\Delta v_{\rm t}$}\\

\multicolumn{1}{c}{}&
\multicolumn{1}{c}{$+100$~K}&
\multicolumn{1}{c}{$+0.3$}&
\multicolumn{1}{c|}{$+0.3 {\rm km/s}$} &
\multicolumn{1}{c}{}&
\multicolumn{1}{c}{$+100$~K}&
\multicolumn{1}{c}{$+0.3$}&
\multicolumn{1}{c}{$+0.3 {\rm km/s}$} \\
\tablerule
C(C$_2$)	&  -0.01  &  0.01   &  0.10 	& Fe\,{\sc ii}	&  -0.07  &  0.15  &  -0.03 \\
N(CN)		&  0.07   &  0.00   &  0.09 	& Co\,{\sc i}	&  0.08	  &  0.05  &  -0.03 \\
\rlap{O([O\,{\sc i}])} &  0.01   &  0.01   &  0.14 	& Ni\,{\sc i}	&  0.04   &  0.05  &  -0.07 \\
Na\,{\sc i}	&  0.15   &  -0.01  &  -0.01 	& Y\,{\sc ii}	&  0.00   &  0.12  &  -0.01 \\
Si\,{\sc i}	&  -0.02  &  -0.06  &  -0.02 	& Zr\,{\sc i}	&  0.17	  &  0.00  &  -0.02 \\
Ca\,{\sc i}	&  0.09   &  0.00   &  -0.05 	& La\,{\sc ii}  &  0.01	  &  0.12  &  -0.01 \\
Sc\,{\sc ii}	&  -0.01  &  0.12   &  -0.09 	& Ce\,{\sc ii} 	&  0.01	  &  0.12  &  -0.01 \\
Ti\,{\sc i }	&  0.12   &  0.00   &  -0.07 	& Pr\,{\sc ii} 	&  0.02	  &  0.13  &  -0.01 \\
V\,{\sc i}	&  0.15   &  0.01   &  -0.07 	& Nd\,{\sc ii}	&  0.01	  &  0.12  &  -0.01 \\
Cr\,{\sc i}	&  0.09   &  0.00   &  -0.08 	& Eu\,{\sc ii}	&  -0.01  &  0.13  &   0.00 \\
Fe\,{\sc i}	&  0.06   &  0.03   &  -0.04 	&  &  &  & \lstrut	\\
C/N		&  -0.28 \hstrut &  0.06	  &  0.06  & \rlap{\textsuperscript{12}C/\textsuperscript{13}C} &
 0.0 0& 0.00 & 1.00  \\
\tablerule
\end{tabular}
\end{footnotesize}
}
\end{center}
\end{table}

\sectionb{3}{RESULTS AND DISCUSSION}

\subsectionb{3.1}{Atmospheric parameters and the iron abundance}

As a result, for $\lambda$~And we have determined the following
atmospheric parameters:  $T_{\rm eff}=4830$~K, log\,$g=2.8$, $v_{\rm
t}=1.6~{\rm km}\,s^{-1}$, [Fe/H] = --0.53, [C/Fe] = 0.09, [N/Fe] = 0.35,
[O/Fe] = 0.45, C/N = 2.21, $^{12}$C/$^{13}$C = 14.  The element
abundances [A/H] and $\sigma$ (the line-to-line scatter) are listed in
Table~3 and compared with results of other investigations in Figure~7.

\vspace{1mm}
\begin{table}[!th]

\begin{center}
\vbox{\small
\parbox{115mm}{\baselineskip=9pt
{\small \bf \ \ Table 3.}{\small \ Element abundances relative to
hydrogen [A/H]. $\sigma$ is a standard deviation in the
mean value determined from the line-to-line scatter within the species.
$N$ is the number of lines used for the abundance determination. }}
\tabcolsep=10pt
\begin{footnotesize}
\begin{tabular}{lrrc|lrrc}
\tablerule
\multicolumn{1}{l}{Element}&
\multicolumn{1}{c}{$N$}&
\multicolumn{1}{c}{[El/H]}&
\multicolumn{1}{c|}{$\sigma$}&
\multicolumn{1}{l}{Element}&
\multicolumn{1}{c}{$N$}&
\multicolumn{1}{c}{[El/H]}&
\multicolumn{1}{c}{$\rm \sigma$} \\
\tablerule
C(C$_2$)         & 1 & $-0.44$& $-$  & Fe\,{\sc ii} & 8 & $-0.54$& 0.07 \\
N(CN)            & 5 & $-0.18$& 0.04 & Co\,{\sc i} & 4 & $-0.43$& 0.07 \\
O([O\,{\sc i}])  & 1 & $-0.08$& $-$  & Ni\,{\sc i} & 19 & $-0.54$& 0.04  \\
Na\,{\sc i}      & 1 & $-0.22$& $-$  & Y\,{\sc ii} & 1 & $-0.38$& $-$  \\
Si\,{\sc i}      & 2 & $-0.31$& 0.03 & Zr\,{\sc i} & 3 & $-0.17$& 0.11 \\
Ca\,{\sc i}      & 3 & $-0.10$& 0.05 & La\,{\sc ii} & 1 & $-0.37$ & $-$  \\
Sc\,{\sc ii}     & 2 & $-0.26$& 0.05 & Ce\,{\sc ii}  & 2 & $-0.51$ & 0.03  \\
Ti\,{\sc i}      & 16& $-0.01$& 0.04 & Pr\,{\sc ii}  & 1 & $-0.15$ & $-$  \\
V\,{\sc i}       & 15& $-0.04$& 0.05 & Nd\,{\sc ii} & 1 & $-0.15$ & $-$ \\
Cr\,{\sc i}      & 5 & $-0.37$& 0.07 & Eu\,{\sc ii} & 1 & $-0.15$& $-$ \\
Fe\,{\sc i}      & 44& $-0.53$& 0.03 &	&	& &\lstrut  \\
\tablerule
\end{tabular}
\end{footnotesize}
}
\end{center}
\end{table}

$\lambda$~And has a long history of high resolution spectral
investigations.  The first were Helfer \& Wallerstein (1968) who
studied
the star by the differential curve-of-growth method.  The chemical
composition of $\lambda$~And by modern methods using stellar atmosphere
models was investigated by McWilliam (1990), Tautvai\v{s}ien\.{e} et
al.\ (1992), Savanov \& Berdyugina (1994), Donati et al.\ (1995).  The
lithium abundance in $\lambda$~And was investigated by Randich et al.\
(1994) and Mallik (1998) and just the main atmospheric parameters
recently were determined by Soubiran et al.\ (2008).

The available values of the effective temperature for $\lambda$~And
cover a wide interval, from 4032~K (Helfer \& Wallerstein 1968)
to 4850~K (Randich et al.\ 1994).  Our result (4830~K) is in good
agreement with the Randich et al.\ determination.  The temperature
determined by Donati et al.  (1995) is by 80~K lower, but this does not
exceed the temperature determination errors.

The values of surface gravities (log~$g$) for $\lambda$~And are in the
interval from 2.0 (Tautvai\v{s}ien\.{e} et al.\ 1992) to 3.1 (McWilliam
1990; Mallik 1998; Soubiran et al.\ 2008).  Our present value
(log~$g=2.8$) is in the best agreement with that of Randich et al.
(1994); the differences with the results of other studies are also
within the error box.

The [Fe/H] values, received for $\lambda$~And, extend from $-0.43$
(Mallik 1998) to $-0.80$ (Tautvai\v{s}ien\.{e} et al.\ 1992).  The
average value of [Fe/H] of the previous studies is $-0.58$.  Our value,
[Fe/H] = --0.53, is close to this average.

\vskip2mm

\begin{figure}[!th]
\vbox{
\centerline{\psfig{figure=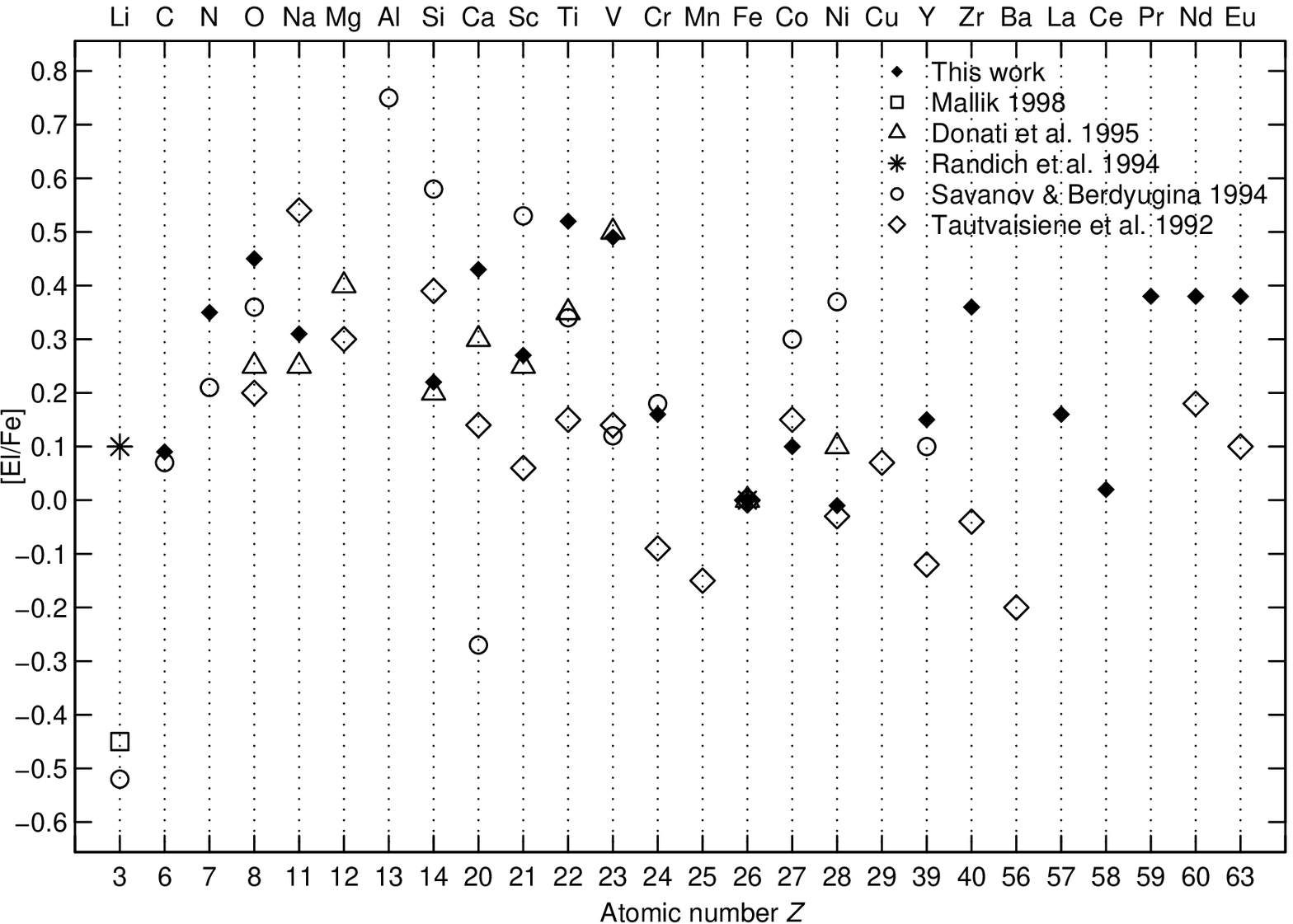,width=120mm,angle=0,clip=}}
\vskip1mm
\captionb{7}{Abundances of elements for $\lambda$~And, as
determined in this work (filled diamonds), Donati et al.\ (1995,
triangles), Savanov \& Berdyugina (1994, circles),
Tautvai\v{s}ien\.{e} et al.\ (1992, empty diamonds). The [Li/Fe] value
from Randich et al.\ (1994) is shown as an asterisk and from
Mallik (1998) as a square.}
}
\end{figure}

\subsectionb{3.2}{Carbon and nitrogen}

For the interpretation of carbon and nitrogen abundances in
$\lambda$~And, we have to remind several evolutionary episodes of
low-mass red giants.

The first opportunity for low-mass stars to modify their surface carbon
and nitrogen abundances happens on their way to the red giant branch
when they undergo the so-called first dredge-up (Iben 1965).  During
this evolutionary stage, the deepening convective envelope mixes the
outer layers of the red giant with the internal matter which has been
CN-processed while the star resided on the main sequence.  Convective
mixing induces a change of the carbon and nitrogen surface abundances.
The atmospheric abundance of $^{12}{\rm C}$ decreases, while the
$^{13}{\rm C}$ and $^{14}{\rm N}$ abundances increase.

However, the classical stellar evolution theory considers stars as
non-rotating and non-magnetic bodies, and the convection is accepted as
the only mixing process.  Observations of CNO elements in a large number
of evolved low-mass giants show much lower C/N and $^{12}{\rm
C}/^{13}{\rm C}$ ratios than the post-dredge-up ratios predicted in the
framework of standard stellar theory (see reviews by Chanam\'{e} et al.\
2005 and Charbonnel 2006).  Therefore, it was concluded that the next
distinct mixing episode in low-mass stars occurs at the so-called red
giant
branch `bump'.  It was surmised that at this evolutionary step the mean
molecular weight gradient, produced by the first dredge-up and
inhibiting mixing processes, is erased by the outwardly-burning hydrogen
shell and results in extra mixing of the convective zone material with
regions hot enough to convert $^{12}{\rm C}$ to $^{13}{\rm C}$.  This
evolutionary phase is referred to as the `bump' of the luminosity
function on the HR diagram and it corresponds to a temporary decrease in
the luminosity and a small increase in the effective temperature of the
star when the chemical discontinuity is removed.

In Figure~8 we compare C/N and $^{12}{\rm C}/^{13}{\rm C}$ ratios of
$\lambda$~And with two theoretical models of extra-mixing.  The first
model, called as `cool bottom processing' (CBP), was calculated by
Boothroyd \& Sackmann (1999).  It includes the deep circulation mixing
below the base of the standard convective envelope and the consequent
processing of CNO isotopes.

\vskip3mm

\vbox{
\centerline{\psfig{figure=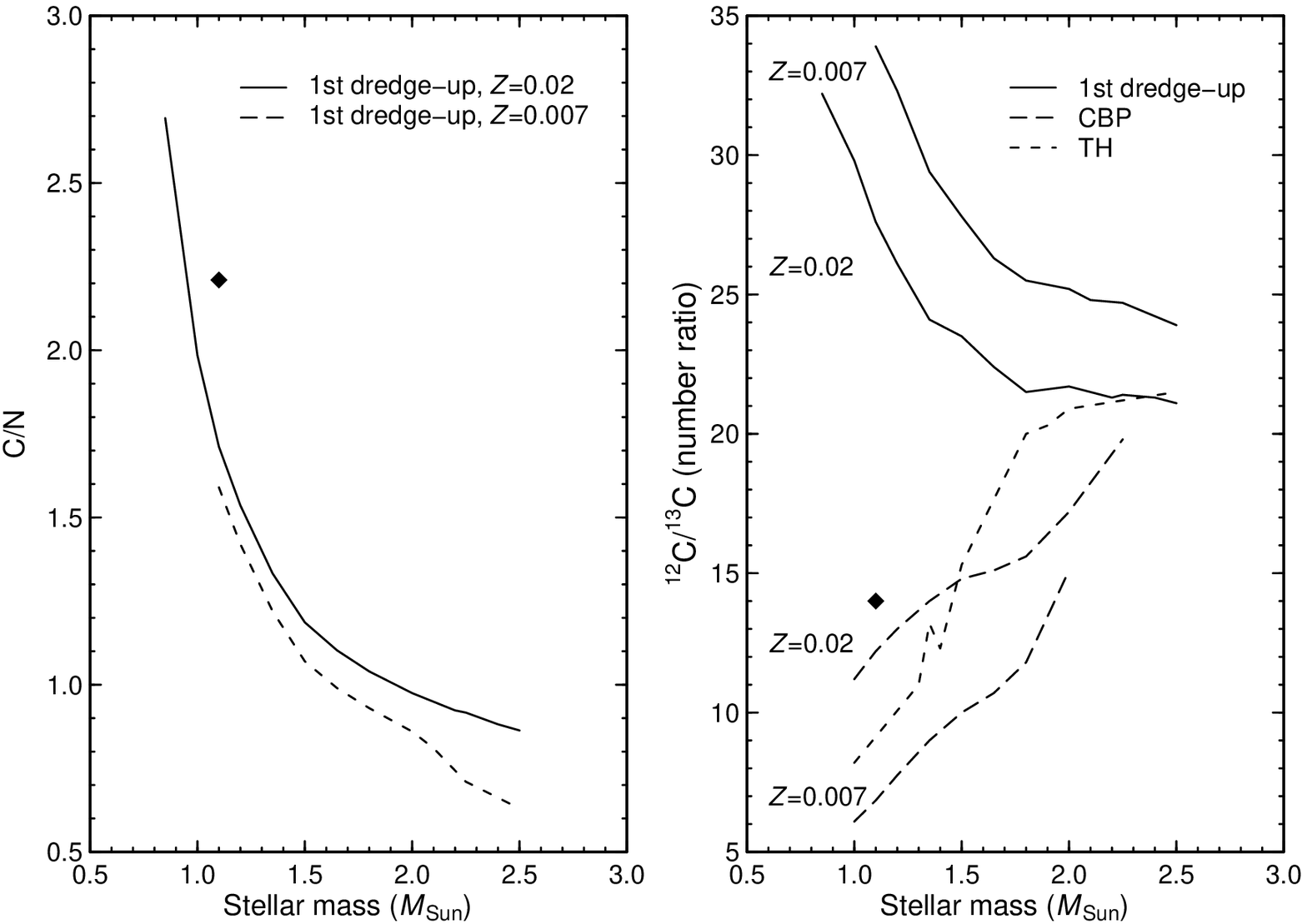,width=120mm,angle=0,clip=}}
\captionb{8}{Comparisons of C/N and $^{12}$C/$^{13}$C ratios of
$\lambda$~And (diamonds) with the theoretical predictions for two
metallicities.  Left panel:  the C/N values in the first dredge-up
(Boothroyd \& Sackman 1999).  Right panel:  the $^{12}$C/$^{13}$C values
for the first
dredge-up and the cool bottom processing (CBP) model by Boothroyd \&
Sackman
(1999) and the model of thermohaline mixing (TH) by Charbonnel \&
Lagarde (2010).}
}

\vskip3mm

Another model, called as the `thermohaline mixing' (TH), was recently
presented by Charbonnel \& Lagarde (2010).  Following Eggleton et al.\
(2006) and Charbonnel \& Zahn (2007), this model considers a double
diffusive instability referred to as thermohaline convection, which had
been discussed long ago in the literature (Stern 1960), as an important
process in evolution of red giants.  This mixing connects the convective
envelope with the external wing of hydrogen burning shell and induces
surface abundance modifications in red giant stars.  The mean molecular
weight inversion at the red giant bump is produced by the reaction
$^3{\rm He(}^3{\rm He,2p)}^4{\rm He}$, as predicted by Ulrich (1972).
According to the authors, the thermohaline mixing does not occur earlier
than the bump, since the magnitude of the mean molecular wight inversion
is small compared to a stabilizing mean molecular weight stratification.

According to the comparison with stellar evolutionary sequences in the
luminosity versus effective temperature diagram by Girardi et al.\
(2000), $\lambda$~And with its luminosity ${\rm
log}\,(L/L_{\odot})=1.37$ is a first ascent giant lying slightly below
the red giant sequence bump indicated at ${\rm log}\,(L/L_{\odot})=1.6$
(Charbonnel \& Lagarde 2010).  According to the mentioned models of
mixing, carbon and nitrogen abundances should be altered only by the
first dredge-up.

However, the position of $\lambda$~And in the $^{12}$C/$^{13}$C versus
stellar mass diagram (Figure~8) indicates that its carbon isotope ratio
is altered by extra mixing.  The low value of the $^{12}$C/$^{13}$C
ratio in $\lambda$~And gives a hint that extra-mixing processes may
start acting in low-mass chromospherically active stars below the
currently predicted place of the bump of the luminosity function of red
giants.  It is interesting that the C/N ratio of $\lambda$~And is
similar to that of II~Peg, another very active RS~CVn-type star with low
metallicity (${\rm [Fe/H]}=-0.4$, ${\rm C/N}\sim 2$, Berdyugina et al.\
1998a).  However, the $^{12}$C/$^{13}$C ratio of II~Peg is not known,
and this prevents evaluation whether this star has undergone additional
mixing.  This is an important task for future.

\subsectionb{3.3}{Other heavy elements}

Abundance ratios of iron peak elements investigated in this work
(Cr, Co, Ni and Sc) to the abundance of Fe are close to the solar ratio.
Only vanadium is overabundant by [V/Fe]\,=\,0.5.  The same value was
obtained by Donati et al.\ (1995); most probably the overabundance is
caused by the hyperfine splitting effects.  However, Savanov \&
Berdyugina (1994) and Tautvai\v{s}ien\,{e} et al.\ (1992) have received
almost solar ratios.  Donati et al.  (1995), applying 11 spectrograms of
$\lambda$~And taken in 1991 and 1994, investigated the dependence of
equivalent widths of lines on the rotational phase and the surface spot
coverage.  They have noticed that for some lines, e.g., Fe\,{\sc i},
the equivalent widths remain almost unchanged in the presence of cool
spots.  However, the lines of V\,{\sc i} and Ti\,{\sc i} increase
noticeably.  Our results show that [El/Fe] ratios are the largest for
these particular elements.

We obtained the following average values of abundances for $\alpha$-, s-
and r-elements:  [$\alpha/{\rm Fe]}=0.4$, ${\rm [s/Fe]}=0.2$ and
${\rm [r/Fe]}=0.4$.  These ratios are by 0.2--0.25 dex lower than
the those of the theoretical models calculated for the thin disk of the
Milky Way by Pagel \& Tautvai\v{s}ien\.{e} (1995, 1997) at the
metallicity $-0.5$~dex.  Similar overabundances of $\alpha$-elements
were obtained in a sample of other RS~CVn stars investigated by Morel et
al.\ (2004).  From the s- and r-process elements, only barium has been
investigated by Morel et al.\ and found to be also overabundant.

Abundances of Mg and Al were not investigated in our work.  As it is
seen from Figure~7, the available [Mg/Fe] values (Donati et al.\ 1995;
Tautvai\v{s}ien\.{e} et al.\ 1992) are in agreement with other
$\alpha$-elements.  Aluminum was investigated only by Savanov \&
Berdyugina (1994) and found to be overabundant even by 0.7~dex.  We also
did not investigate the abundance of lithium.  The available [Li/Fe]
values for $\lambda$~And are quite different.  Savanov \& Berdyugina
(1994) and Mallik (1998) find about $-0.5$~dex, while the result of
Randich et al.\ (1994) is $+0.1$~dex.

More observational and theoretical studies of RS CVn stars are essential
in trying to answer the questions of the role of magnetic fields in
stellar plasma dynamics, the relation between the stellar structure and
energy balance, the interaction between stellar rotation and orbital
motion.

\thanks{ This project was supported by the European Commission
through the Baltic Grid project.}

\References

\parskip=-0.1pt

\refb Alekseev~I.~Yu., Kozhevnikova~A.~V. 2005, Astrophysics, 48, 450

\refb Alonso~A., Arribas~S., Mart\'{i}nez-Roger~C. 1999, A\&AS, 140, 261

\refb Anstee~S.~D., O'Mara B.J. 1995, MNRAS, 276, 859

\refb Audard~M.,~G\"{u}del M., Sres A., Mewe R., Raassen A. A. J., Behar
E., Foley C. R., van der Meer R. L. J. 2001, in {\it Stellar Coronae in
the Chandra and XMM-Newton Era}, eds.  F. Favata \& J. J. Drake, ASP
Conf.  Ser., 277, 65

\refb Audard~M.,~G\"{u}del M., Sres A., Mewe R., Raassen A.\,A.\,J.,
Behar E., Mewe R. 2003, A\&A, 398, 1137

\refb Barklem~P.~S., O'Mara B.J. 1997, MNRAS, 290, 102

\refb Barklem~P.~S., O'Mara B.J., Ross J.E. 1998, MNRAS, 296, 1057

\refb Berdyugina S.~V. 2005, Living Rev.  Solar Phys., 2, 8

\refb Berdyugina S.~V., Jankov S., Ilyin I., Tuominen I., Fekel F.~C.
1998a, A\&A, 334, 863

\refb Berdyugina S.~V., Berdyugin A.~V., Tuominen I., Fekel F.~C. 1998b,
A\&A, 340, 437

\refb Berdyugina S.~V., Ilyin I., Tuominen I. 1999, A\&A, 347, 932

\refb Berdyugina S.~V., Berdyugin A.~V., Ilyin I., Tuominen I. 2000,
A\&A, 360, 272

\refb Biehl~D. 1976, Diplomarbeit, Christian-Albrechts-Universit\"at
Kiel, Institut f\"ur Theoretishe Physik und Sternwarte

\refb Boothroyd~A.~I., Sackman~I.~J. 1999, ApJ, 510, 232

\refb Bopp~B.~W., Noah P. V. 1980, PASP, 92, 717

\refb Bowyer~S., Drake J.~J., Vennes S. 2000, ARA\&A, 38, 231

\refb Calder~W.  A. 1938, Bull.  Harvard Obs., 907, 20

\refb Chanam\'{e} J., Pinsonneault M., Terndrup D. 2005, ApJ, 631, 540

\refb Charbonnel C. 2006, in {\it Stars and Nuclei:  Tribute to Manuel
Forestini}, eds.  T. Montmerle \& C. Kahane, EAS Publ.  Ser., 19, 125

\refb Charbonnel C., Lagarde N. 2010, A\&A, accepted for publication
(arXiv1006.5359)

\refb Charbonnel C., Zahn J.-P. 2007, A\&A, 467, L15

\refb Chisari~D.,  Lacona~G. 1965, Mem. Soc. Astron. Ital., 36, 463

\refb De~Medeiros~J.~R., Da~Silva~J.\,R.\,P., Maia~M.\,R.\,G. 2002, ApJ,
578, 943

\refb Donati J.-F., Henry G. W., Hall D. S. 1995, A\&A, 293, 107

\refb Drake J. J. 1996, in {\it Cool Stars, Stellar Systems and the
Sun}, 9th Cambridge Workshop, eds.  R. Pallavicini \& A. K. Dupree, ASP
Conf.  Ser., p.\,203

\refb Drake J. J. 2002, in {\it Stellar Coronae in the Chandra and
XMM-Newton Era}, eds.  F. Favata \& J. J. Drake, ASP Conf.  Ser., 277,
75

\refb Eggleton P. P., Dearborn D.\,S.\,P., Lattanzio J. C. 2006,
Science, 314, 1580

\refb Frasca A., Biazzo K., Tas G., Evren S., Lanzafame A. C. 2008,
A\&A, 479, 557

\refb Girardi L., Bressan A., Bertelli G., Chiosi C. 2000, A\&AS, 141,
371

\refb Gondoin~P. 2007, A\&A, 464, 1101

\refb Gonzalez G., Lambert D. L., Wallerstein G. et al. 1998, ApJS, 114,
133

\refb Grevesse N., Sauval A. J. 2000, in {\it Origin of Elements in the
Solar System, Implications of Post-1957 Observations}, ed. O. Manuel,
Kluwer, p.\,261

\refb Gurtovenko~E.~A., Kostik~R. I. 1989, in {\it Fraunhofers
spectrum and a System of Solar Oscillator Strengths}, Naukova Dumka,
Kyiv, Ukraine

\refb Gustafsson~B., Edvardsson~B., Eriksson~K., J\o{}rgensen~U.~G.,
Nordlund~\AA., Plez~B. 2008, A\&A, 486, 951

\refb Hakkila~J., Myers~J.~M., Stidham~B.~J., Hartmann~D.~H. 1997, AJ,
114, 2043

\refb Hall~D.~S. 1976, in {\it Structure and Evolution of Close Binary
Systems}, eds.  P. Eggleton, S. Mitton \& J. Whelan, IAU Symp. 73,
p.\,381

\refb Hauck~B., Mermilliod~M. 1998, A\&AS, 129, 431

\refb Helfer~H.  L., Wallerstein~G. 1968, ApJS, 16, 1

\refb Iben I. 1965, ApJ, 142, 1447

\refb Ilyin~I.~V. 2000, in {\it High Resolution SOFIN CCD Echelle
Spectroscopy}, PhD thesis, University of Oulu, Finland, 266 p.

\refb Jeffers~S.~V. 2005, MNRAS, 359, 729

\refb Johansson~S., Litzen~U., Lundberg~H., Zhang~Z. 2003, ApJ, 584,
107

\refb Jordan C., Doschek G. A., Drake J. J., Galvin A. B., Raymond J. C.
1998, in {\it The Tenth Cambridge Workshop on Cool Stars, Stellar
Systems and the Sun}, ASP Conf.  Ser. 154, 91

\refb Katz~D., Favata~F., Aigrain~S., Micela~G. 2003, A\&A, 397, 747

\refb Kurucz~R.L. 2005, {\it New Atlases for Solar Flux, Irradiance,
Central Intensity and Limb Intensity}, Mem. Soc.  Astron.  Ital.
Suppl., 8, 189

\refb Lawler J. E., Wickliffe M. E., Den Hartog E. A. 2001, ApJ, 563,
1075

\refb Mallik S. V. 1998, A\&A, 338, 623

\refb M\"{a}ckle R., Holweger H., Griffin R., Griffin R. 1975, A\&A, 38,
239

\refb McWilliam A. 1990, ApJS, 74, 1075

\refb Mirtorabi M. T., Wasatonic R., Guinan E. F. 2003, AJ, 125, 3265

\refb Montesinos B., Gim\'{e}nez A., Fern\'{a}ndez-Figueroa M. J. 1988,
MNRAS, 232, 361

\refb Morel~T., Micela~G., Favata~F., Katz~D., Pillitteri~I. 2003, A\&A,
412, 495

\refb Morel~T., Micela~G., Favata~F., Katz~D., Pillitteri~I. 2004, A\&A,
426, 1007

\refb Olivier J. P. 1974, PASP, 87, 695

\refb O'Neal~D., Neff~J.~E., Saar~S.~H. 1998, ApJ, 507, 919

\refb Padmakar, Pandey S.~K. 1999, A\&AS, 138, 203

\refb Pagel B.\,E.\,J., Tautvai\v{s}ien\.{e} G. 1995, MNRAS, 276, 505

\refb Pagel B.\,E.\,J., Tautvai\v{s}ien\.{e} G. 1997, MNRAS, 288, 108

\refb Pallavicini~R., Randich~S., Giampapa~M.S. 1992, A\&A, 253, 185

\refb Piskunov~N.~E., Kupka~F., Ryabchikova~T.~A., Weiss~W.~W.,
Jeffery~C.~S. 1995, A\&A, 112, 525

\refb Poe C. H., Eaton J. A. 1985, ApJ, 289, 644

\refb Randich~S., Gratton~R., Pallavichini R. 1993, A\&A, 273, 194

\refb Randich~S., Giampapa M. S., Pallavichini R. 1994, A\&A, 283, 893

\refb Rodon\'{o}~M. 1965, {\it PhD Thesis}, University of Catania, Italy

\refb Rodon\'{o}~M., Lanza~A.~F., Catalano~S. 1995, A\&A, 301, 75

\refb Sanz-Forcada~J., Favata~F., Micela~G. 2004, A\&A, 416, 281

\refb Savanov~I.~S., Berdyugina~S.~V. 1994, Astronomy Letters, 20, 227

\refb Simmons G. J., Blackwell D. E. 1982, A\&A, 112, 209

\refb Soubiran C., Bienaym\'{e} O., Mishenina T., Kovtyukh V. V. 2008,
A\&A, 480, 91

\refb Stern M. E. 1960, Tellus, 12, 172

\refb Strassmeier K. G., Briguglio R., Granzer T., Tosti G. et al. 2008,
A\&A, 490, 287

\refb Tautvai\v{s}ien\.{e}~G., Kriukelis~S., Bikmaev~I. 1992, Baltic
Astronomy, 1, 450

\refb Uesugi~A., Fukuda~I. 1982, {\it Catalogue of Stellar Rotational
Velocities}, Kyoto University

\refb Ulrich R. K. 1972, ApJ, 172, 165

\refb Uns\"{o}ld A. 1955, {\it Physik der Stern Atmosph\"{a}ren}, Zweite
Auflage, Springer-Verlag, Berlin

\refb van Leeuwen~F. 2007, {\it Hipparcos, the New Reduction of the Raw
Data}, Astrophysics and Space Science Library, Vol. 350

\refb Walker E. C. 1944, JRASC, 38, 249

\refb Zahn J.-P. 1977, A\&A, 57, 383

\end{document}